\def\bea{\begin{eqnarray}}
\def\eea{\end{eqnarray}}

\def\beq{\begin{equation}}
\def\eeq{\end{equation}}

\def\squareforqed{\hbox{\rlap{$\sqcap$}$\sqcup$}}
\def\qed{\ifmmode\squareforqed\else{\unskip\nobreak\hfil
\penalty50\hskip1em\null\nobreak\hfil\squareforqed
\parfillskip=0pt\finalhyphendemerits=0\endgraf}\fi}

\def\<{\langle}
\def\>{\rangle}

\documentstyle[prb,twocolumn,aps,rotate,epsf,floats,amssymb,amsmath]{revtex}

\newcommand{\bz}{{\boldsymbol z}}

\newcommand{\bH}{{\boldsymbol H}}

\newcommand{\bOmega}{{\boldsymbol \Omega}}
\newcommand{\obOmega}{{\overline{\boldsymbol \Omega}}}
\newcommand{\bDelta}{{{\boldsymbol \Delta}}}

\newcommand{\ov}{{\overline{v}}}

\begin{document}
\setlength{\unitlength}{1cm}
\renewcommand{\arraystretch}{1.4}

\title{Low temperature electronic properties of Sr$_2$RuO$_4$ III: \\
       Magnetic fields}  

\author{Ralph Werner}
\address{Physics Department, Brookhaven National
Laboratory, Upton, NY 11973-5000, USA and \\ Institut f\"ur Theorie der 
Kondensierten Materie, Universit\"at Karlsruhe, 76128 Karlsruhe, Germany}  

\date{\today}

\maketitle

\centerline{Typeset using REV\TeX}

\begin{abstract} 
Based on the microscopic model introduced previously the observed
specific heat and  ac-susceptibility data in the superconducting phase
in Sr$_2$RuO$_4$ with applied magnetic fields are described
consistently within a phenomenological approach. Discussed in detail
are the temperature dependence of the upper critical fields $H_{\rm
c2}$ and $H_{\rm 2}$, the dependence of the upper critical fields on 
the field direction, the linear specific heat below the
superconducting phase transition as a function of field or
temperature, the anisotropy of the two spatial components of the order
parameter, and the fluctuation field $H_{\rm p}$. 
\end{abstract}
\pacs{PACS numbers: 63.20.Kr, 75.10 Jm, 75.25 +z}


\section{Introduction}

The discovery of superconductivity below $T_c\sim 1$ K in
Sr$_2$RuO$_4$ has quickly triggered a large amount of
interest\cite{MHY+94} because of the unconventional
properties\cite{MRS01} and the initially proposed analogy\cite{RS95}
to $^3$He. The material is tetra\-gonal at all
temperatures.\cite{BRN+98} The three bands cutting the Fermi level 
with quasi two-dimensional Fermi surfaces\cite{Oguc95,MJD+96} can be
mainly associated with the three $t_{2g}$ orbitals of the Ru$^{4+}$
ions.\cite{SCB+96,MRS01} The transport is Fermi liquid
like\cite{MYH+97,MIM+98,IMK+00} for $T_c < T < 30$ K and strongly
anisotropic along the $c$ axis.\cite{MHY+94} The enhanced specific
heat, magnetic susceptibility and electronic mass indicate the
presence of significant correlations.\cite{MHY+94,MJD+96,PSKT98} The
specific heat,\cite{NMF+98,NMM99,NMM00} thermal
conductivity,\cite{STK+02} and nuclear quadrupole resonance
(NQR)\cite{IMK+00} are consistent with two-dimensional gapless
fluctuations in the superconducting phase. For a more detailed
overview see Refs.\ \onlinecite{MRS01,Wern02a}.  

The present work is the last part of a series of three. In part I (Ref.\
\onlinecite{Wern02a}) the quasi one-dimensionality of the kinetic energy
of the $d_{zx}$ and $d_{yz}$ electrons has been used to derive an
effective microscopic model. At intermediate coupling the interaction
leads to a quasi one-dimensional model for the magnetic degrees of
freedom and two-dimensional correlations in the charge sector. The
normal phase properties are described consistently.
In part II (Ref.\ \onlinecite{Wern02b}) it is shown that the
pair-correlations are enhanced by inter-plane umklapp-scattering as a
consequence of the body centered crystal structure. The inter-plane
coupling can be treated mean-field like.\cite{ABG99} The order
parameter is a spin triplet with two slightly anisotropic spatial
(flavor) components. The model consistently can account for the
experimental data concerning the temperature dependence of the pair
density, specific heat, NQR and muon spin relaxation ($\mu$SR) times
and susceptibility. The crucial difference to previous
approaches\cite{RS95,Bask96,Sigr00,Agte01} is that the relevant internal
degrees of freedom of the superconducting order parameter have been
extracted from the material specific microscopic model.

The present paper is based on a phenomenological model that is
consistent with the results derived in Refs.\ \onlinecite{Wern02a} and
\onlinecite{Wern02b}. The upper critical
field\cite{NMM99,YMNF96,MMN+00,YAM+01} can be described by mean-field
theory. The specific heat data\cite{NMM99} are in quantitative
agreement with the presence of a cubic term in the Landau expansion of
the free energy (Sec.\ \ref{sectionmf}). The various observed critical
fields as well as their dependence on the direction of the applied 
field\cite{YMNF96,MMN+00,YAM+01} are well described within the
framework of the flavor degrees of freedom of the model (Sec.\
\ref{sectionH}).


\section{Mean-field theory}\label{sectionmf}

In Ref.\ \onlinecite{Wern02b} it has been shown that the
thermodynamic properties of Sr$_2$RuO$_4$ near the superconducting
phase transition are well described by a Landau expansion in the
pair excitation energy gap $\Delta$.
\begin{equation}\label{FLandau}
F_{\Delta}  =  F_{0} - A\  t\ \Delta^2 + D\ \Delta^3  + B\ \Delta ^4 +
     {\rm O}(\Delta^5)
\end{equation}
Here $t=1-T/T_c \ll 1$ is the reduced temperature. The presence of the
term $\sim \Delta^3$ can be motivated by effectively integrating out
the two-dimensional, gapless order parameter
fluctuations.\cite{Wern02b} Gauge invariance is preserved
since $\Delta \propto |e^{i\phi_{\rm G}}\langle P \rangle |$ for pair
operators $P$ and arbitrary gauge fields $\phi_{\rm G}$. To be 
consistent with the notation of Ref.\ \onlinecite{Wern02b} the
coefficients are given as  
$
A = N \left(64\ V_{0}\right)^{-1}$,
$
B = N \frac{b}{(8\ V_{0})^4}$,
$
D = N \frac{s_0^2}{128\ \ov^2_{\rm eff}}$.
The parameters define the number of Ru ions $N$, the effective
velocity of the elementary electronic excitations above the gap
$\ov_{\rm eff}\approx 22$ K, the numerical prefactor $s_0 \approx 4.7$
K, and the inter-plane pair hopping potential $V_{0}\approx 6$
K. If one assumes that the excitations of the magnetic degrees of
freedom of the order paramter are gapped through spin-flavor coupling
the numbers are $\ov_{\rm eff}\approx 38$ K and $s_0 \approx 2.7$. 

Comparing with experiments\cite{LFK+98,LGL+00} the superconducting gap
has been found to be linear in the reduced temperature\cite{Wern02b}
over a rather large interval, i.e.,
\begin{equation}\label{DeltaofT}
\Delta(T)\big|_{0 \le t < 0.5} \approx \frac{2 A}{3 D}\ t 
                         \approx 0.8\ V_{0}\ t\,.
\end{equation}
Since $\Delta\ge 0$ the phase transition described by Eqs.\
(\ref{FLandau}) and (\ref{DeltaofT}) is of third order in the sense of
Ehrenfest's definition.\cite{Ehre33}


\subsection{Upper critical fields $\bH_{\rm c2}$}\label{sectionHcm}

\begin{figure}[bt]
   \epsfxsize=0.48\textwidth
   \centerline{\epsffile{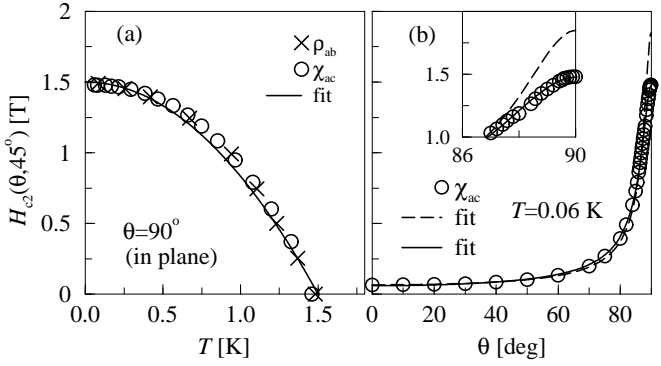}}
   \centerline{\parbox{\textwidth}{\caption{\label{Hcps}
   \sl Upper critical magnetic fields. Circles are from
   ac-susceptibility measurements Ref. \protect\onlinecite{YAM+01}, 
   crosses are from resistivity measurements Ref.\
   \protect\onlinecite{ANM99}. (a) $H_{\rm cm}$ along [110] as a
   function of temperature, fit from Eq.\ (\protect\ref{HcofT}). (b)
   $H_{\rm c2}(\theta,45^\circ)$ as a function of the polar angle
   $\theta$. The dashed line is a fit from Eq.\
   (\protect\ref{Hcoftheta}), the solid line is a fit from Eq.\
   (\protect\ref{Hcofthetaani}). The inset is an enlargement of the
   region $\theta\sim 90^\circ$.}}}        
   \end{figure}

The field above which superconductivity disappears is denoted $H_{\rm
c2}(\theta,\phi)$. The azimuthal $\phi$ and polar $\theta$ angles are
defined as $H(0,\phi) \| [001]$ and $H(90^\circ,0) \| [100]$. The
critical field is maximal along $[110]$,\cite{MMN+00,YAM+01} i.e.,
$H_{\rm cm}(T) = H_{\rm c2}(90^\circ,45^\circ)$. $H_{\rm cm}(T)$ is
plotted in Fig.\ \ref{Hcps}(a) as derived from ac-susceptibility
measurements (circles)\cite{YAM+01} and resistivity data
(crosses).\cite{ANM99} The fit to the phenomenological
curve\cite{MMM99}  
\begin{equation}\label{HcofT}
\frac{H_{\rm cm}(T)}{H_{\rm cm}(T=0)} =  1 - 
                    \left(\frac{T}{T_c(H=0)}\right)^2
\end{equation}
with $H_{\rm cm}(T=0)=1.5$ T and $T_c(H=0)=1.5$ K reproduces the
experimental data satisfactorily [full line in Fig.\
\ref{Hcps}(a)]. The deviation of Eq.\ (\ref{HcofT}) from the BCS
predictions $\lim_{T\to 0} \frac{H_{\rm cm}(T)}{H_{\rm cm}(T=0)}
\approx 1 - 1.06 \frac{T^2}{T^2_c(H=0)}$ and $\lim_{T\to T_c}
\frac{H_{\rm cm}(T)}{H_{\rm cm}(T=0)} \approx 1.74  [1 -
\frac{T}{T_c(H=0)}]$ is smaller than the experimental uncertainty.  

The experimental uncertainties in the determination of $H_{\rm cm}$ lie
in part in the ambiguities between $T_{\rm c}$ and the ``mid transition''
value $T_{\rm cm}$ due to the linear order parameter as discussed in Ref.\
\onlinecite{Wern02b}. In Ref.\ \onlinecite{YAM+01} the possible
violation of the Werthamer-Helfand-Hohenberg (WHH) formula\cite{WHH66}
is discussed. The WHH formula relates $H_{\rm cm}(T=0)$ and the
derivative $d H_{\rm cm}(T)/dT|_{T=T_c}$. Since the experimental
estimate relies solely on the two points closest to $T_c(H=0)$ the
discrepancies must very likely be attributed to experimental 
uncertainties. 

Figure \ref{Hcps}(b) shows the critical field $H_{\rm
c2}(\theta,45^\circ)$ as a function of the polar angle. The fit 
from the Landau-Ginzburg anisotropic effective mass 
approximation\cite{Tink96} defined by 
\begin{equation}\label{Hcoftheta}
H_{\rm c2}(\theta,45^\circ) =
\frac{H_{\rm c2}(\theta=0,45^\circ)} 
      {|\cos\theta| \sqrt{1 + \tan^2\theta/R^2_m}}
\end{equation}
gives an effective mass ratio of $R_m\approx 28$ [dashed line in Fig.\
\ref{Hcps}(b)]. 

While the experimental data\cite{YAM+01} are well reproduced for 
$\theta < 87^\circ$, the fit function overestimates the in-plane critical
field at $\theta=90^\circ$ by roughly 20\% [inset of Fig.\
\ref{Hcps}(b)]. Such an enhancement of the in-plane coupling to the
magnetic field is expected from the 20\% easy plane anisotropy of the
spin-one Cooper pairs that was implied from the $\mu$SR relaxation
time anisotropy.\cite{Wern02b} The easy plane configuration can be
incorporated into Eq.\ (\ref{Hcoftheta}) by enhancing the in-plane
coupling by a factor $g_\parallel = 1.23$.
\begin{equation}\label{Hcofthetaani}
H_{\rm c2}(\theta,45^\circ) =
\left|\left(\begin{array}{c}
g_\parallel \sin\theta \\ \cos\theta
\end{array}\right)\right|\
\frac{H_{\rm c2}(\theta=0,45^\circ)} 
      {|\cos\theta| \sqrt{1 + \tan^2\theta/R^2_m}}
\end{equation}
The experimental data are then well described with $R_m = 20$ [solid
line in Fig.\ \ref{Hcps}(b)]. As expected this value excellently
matches the in-plane to out-of-plane hopping ratio
$R_m=t_0/t_\perp$.\cite{Wern02a} 

Both the temperature dependence of $H_{\rm cm}$ and the angular
dependence of $H_{\rm c2}(\theta,45^\circ)$ strongly support the
applicability of the mean-field approach.


\subsection{Landau theory with magnetic field}\label{sectionFL}

\begin{figure}[bt]
   \epsfxsize=0.48\textwidth
   \centerline{\epsffile{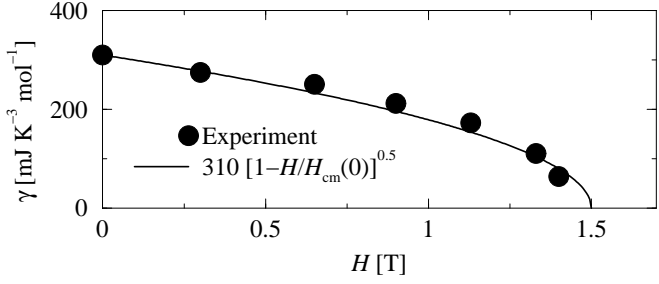}}
   \centerline{\parbox{\textwidth}{\caption{\label{TcHslope}
   \sl Coefficient $\gamma = \frac{8}{9} \frac{A^3}{T_{\rm c}^3 D^2}$
   as determined from the measurements in Ref.\
   \protect\onlinecite{NMM00} as a function of the applied magnetic
   field (symbols). The full line [Eq.\ (\ref{defgamma})] is the fit
   consistent with the mean-field approach.}}} 
   \end{figure}

Using Eqs.\ (\ref{FLandau}) and (\ref{DeltaofT}) the specific heat
just below the superconducting transition becomes
\begin{equation}\label{CsofT}
\frac{C_s|_{t\ll 1}}{T_{\rm c}} = \frac{C_n}{T_{\rm c}} + 
\frac{8}{9}\frac{A^3}{T_{\rm c}^3 D^2}\ (T_{\rm c} - T)
- {\rm O}(t^2)\,.
\end{equation}
The critical temperature depends on the magnetic field $T_{\rm c} =
T_{\rm c}(H)$. The normal phase contribution is constant with
$\frac{C_n}{T_{\rm c}}\approx 37.5\ \frac{\rm mJ}{\rm K^2 mol}$. The
coefficient $\frac{8}{9} \frac{A^3}{T_{\rm c}^3 D^2}$ can
be determined from the measurements in Ref.\ \onlinecite{NMM00} as a
function of the applied magnetic field. The results are plotted in
Fig.\ \ref{TcHslope} (symbols) and can be fitted by 
\begin{equation}\label{defgamma}
\gamma = \frac{8}{9} \frac{A^3}{T_{\rm c}^3 D^2} =
310 \sqrt{1-\frac{H}{H_{\rm cm}(0)}}\ \frac{\rm mJ}{\rm K^3 mol} 
\end{equation}
with $H_{\rm cm}(0)=1.5$ T (full line). 

The reduced temperature at a given temperature is altered as the
magnetic field is changed, i.e., $t=t(H)$, because of the resulting
change in the critical temperature. The latter can be obtained by
inverting Eq.\ (\ref{HcofT}) to give $T_{\rm c}(H)$. Defining the
reduced field $h(T) = 1 - \frac{H}{H_{\rm cm}(T)}$ at a given
temperature and expanding Eq.\ (\ref{HcofT}) for $t\ll 1$ one finds
the relation   
\begin{equation}\label{tofH}
t(H)\ \frac{T_{\rm c}^2(H)}{T_{\rm c}^2(0)}
\approx \frac{1}{2}
\frac{H_{\rm cm}(T)}{H_{\rm cm}(0)}\ h(T)\,
\end{equation}
between the reduced temperature and the reduced field. Using Eq.\
(\ref{tofH}) and the fit of the coefficient $\gamma$ as 
given in Eq.\ (\ref{defgamma}) the specific heat as a function of the
magnetic field becomes at a given temperature for $0\le h \ll 1$
\begin{equation}\label{Csofh}
\frac{C_s|_{h\ll 1}}{T} = \frac{C_n}{T} + 
\frac{155\ \rm mJ}{\rm T K^2 mol}\ (H_{\rm cm}-H) - {\rm O}(h^2)\,.
\end{equation}
From the experimental results in Ref.\ \onlinecite{NMM00} one finds
for $0.4 {\rm K} < T < 1.5 {\rm K}$ values for the linear field
coefficient of $140$ to $160$ $\frac{\rm mJ}{\rm T K^2 mol}$ in
excellent agreement with the prediction. 

For $H>1.4$ T or $T_{\rm c}<0.4$ K corrections to the fit of $\gamma$
must be expected. The data in Fig.\ \ref{TcHslope} are consistent with
a faster drop of $\gamma$ for $H>1.4$ T than anticipated by the fit
from Eq.\ (\ref{defgamma}). At the the same time the region of
validity of Eq.\ (\ref{tofH}) is reduced to extremely small values of
$h$. For $T\to 0$ the term linear in $H$ in Eq.\ (\ref{Csofh}) then
becomes negligible. The higher order terms in Eq.\ (\ref{Csofh}) are
negative and are consistent with the sharp drop of the specific heat
with $h$ for $T\to 0$.\cite{NMM00} 
Detailed specific heat measurements in the range
$1.4\ {\rm T} < H < 1.5\ {\rm T}$ are desirable for further
clarification.

Using the definition Eq.\ (\ref{defgamma}) and Eq.\ (\ref{HcofT}) to
express $T_{\rm c}(H)$ as a function of the magnetic field gives 
\begin{equation}\label{DofH} 
D(H) = 4.88\ {\rm K}\ \sqrt{\frac{A}{T_c(0)}}^3\ 
       \left(1-\frac{H}{H_{\rm cm}(0)}\right)^{-1} .
\end{equation}
Assuming that the coefficient $A$ is independent\cite{Wern02b} of the
field $H$ and replacing $D \to D(H)$ in Eq.\ (\ref{DeltaofT}) one
finds    
\begin{eqnarray}\label{DeltaHofT}
\Delta_H(T)\big|_{t \ll 1} 
&\approx& \Delta_0\ \sqrt{1 - \frac{H}{H_{\rm cm}(0)}}\
   \frac{T_{\rm c}(H) - T}{T_{\rm c}(0)}\,,
\end{eqnarray}
with $\Delta_0 = 4.9\ {\rm K}$.
Applying Eq.\ (\ref{tofH}) yields
\begin{eqnarray}\label{DeltaTofH}
\Delta_T(H)\big|_{h \ll 1} 
&\approx& \Delta_0\ \frac{H_{\rm cm}(T) - H}{H_{\rm cm}(0)}\,.
\end{eqnarray}

Equations (\ref{DofH}) -- (\ref{DeltaTofH}) only are meaningful under
the assumption that the coefficient $A$ is independent of the field
$H$. They yield meaningful results consistent with experiments within
the present approach justifying the assumption made. An unambiguous
quantitative test of Eqs.\ (\ref{DeltaHofT}) and (\ref{DeltaTofH}) can
be obtained via excess current measurements.\cite{Wern02b,LGL+00}


\subsection{Low temperature specific heat}\label{sectionLTSH}

For $T\to 0$ and $H < H_{\rm sg} \approx 0.12$ T the specific heat of
Sr$_2$RuO$_4$ shows a sharp linear increase as a function of the
applied magnetic field.\cite{NMM00} This phenomenon can be easily
understood in terms of a small gap in the magnetic excitation spectrum
of internal magnetic degrees of freedom of the superconducting order
parameter which is induced by spin-flavor
coupling.\cite{Wern02b,Volo92} The applied magnetic field splits the
three components of the spin triplet described by the SO(3)
vector\cite{Wern02b} $\obOmega_{\rm s}$ and the lowest spin state is
decreased in energy until the spin gap is closed. The spin gap can be
estimated as $\Omega_{A} \approx \mu_{\rm B} H_{\rm sg} = 0.08$
K. This value is consistent with an estimate obtained via the
analogy\cite{Spinspectrumquotet} to $^3$He-$A$ and suggests that for
temperatures $T > 0.1$ K the magnetic excitations of the Cooper pair
moments can be considered as gapless.

For $H > 0.12$ T and at low temperatures $0.1\ {\rm K} \le T \ll
T_{\rm c}$ the experimental specific heat can be described by a
contribution linear in $T$ and a contribution quadratic in
$T$.\cite{NMM00} The quadratic contribution stems from the quasi
two-dimensional gapless fluctuations of the internal degrees of freedom
of the order parameter. In the presence of a magnetic field the
magnetic degeneracy of the magnetic pair order parameter components is
lifted by the Zeeman splitting. The non-linear sigma model describing
the gapless fluctuations of the order parameter components is thus
reduced from SO(3)$\otimes$SO(2) to one channel or SO(2) symmetry. The
quadratic specific heat contribution then is [c.f.\ Eq.\ (70) in Ref.\ 
\onlinecite{Wern02b}]    
\begin{equation}\label{Cs(2)}
C^{(2)}_s|_{T \to 0}= \frac{3\,\zeta(2)}{\pi}\ \frac{T^2}{(v_\nu)^2}
\approx 1.15\ \frac{T^2}{(\ov_{\rm eff})^2}\,.
\end{equation}
Equation (\ref{Cs(2)}) predicts a specific heat contribution quadratic
in temperature that is basically independent of the magnetic
field. This result is consistent with experiments (c.f.\ Fig.\ 6 in
Ref.\ \onlinecite{NMM00}) where the coefficient has been determined
for $1.4\ {\rm T} \ge H > 0.12$ T as $C^{(2)}_{s,\rm exp}|_{T \to
0}\approx 44\ \frac{\rm mJ}{\rm K^3 mol}\ T^2$. Comparison with Eq.\
(\ref{Cs(2)}) leads to $\ov_{\rm eff} \approx 15$ K which compares
reasonably well with the value of 22 K determined in the absence of a
magnetic field.\cite{Wern02b} The change of the excitation velocity is
conceivable in a magnetic field as well as quadratic temperature
contributions from the vortex lattice.\cite{KRF+00,HA99}

The linear contribution to the specific heat for $H > 0.12$ T stems
form the excitations of the (non-degenerate) lowest spin-triplet
state discussed above as well as from the vortex lattice.


\section{Critical magnetic fields}\label{sectionH}

Experimentally different critical magnetic fields have been observed
in
Sr$_2$RuO$_4$.\cite{MHY+94,NMM00,YMNF96,MMN+00,YAM+01,TNM+01,STM+02}
So far the physical implications have been discussed controversially.


\subsection{In-plane anisotropy of $H_{\rm c2}$}\label{sectionHc2}

\begin{figure}[bt]
   \epsfxsize=0.48\textwidth
   \centerline{\epsffile{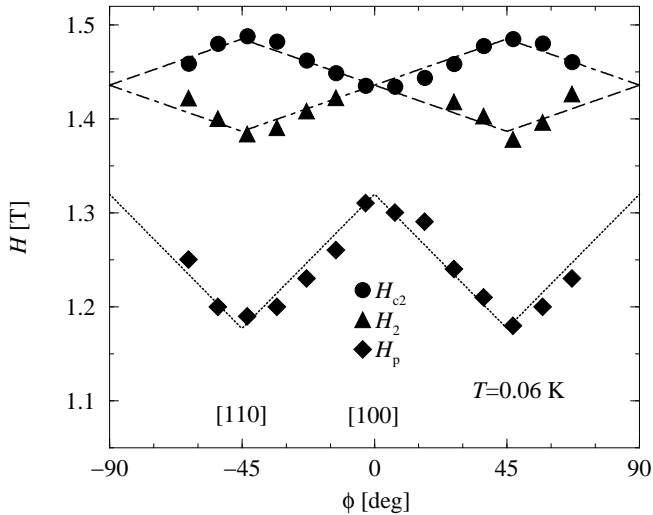}}
   \centerline{\parbox{\textwidth}{\caption{\label{Hcphi}
   \sl In-plane critical magnetic fields. Symbols are from
   ac-susceptibility measurements Ref.\
   \protect\onlinecite{YAM+01}. The broken line is $H_{{\rm
   c},x}(\phi)$, the dash-dotted line $H_{{\rm 
   c},y}(\phi)$ as defined in Eq.\ (\protect\ref{Hcnu}) with $H_{{\rm
   c},u} = 1.44$ T and $H_{{\rm c},a} = 0.05$ T. The dotted line is
   the crossover field $H_{\rm p}$ as defined in Eq.\
   (\protect\ref{Hpphi}) that marks the onset of flavor
   fluctuations.}}}           
   \end{figure}

The ac-susceptibility measurements reveal an angular alternation of the
in-plane upper critical field $H_{\rm c2}(\phi)$ with four fold
symmetry\cite{MMN+00,YAM+01} (full circles in Fig.\ \ref{Hcphi}). A
second transition is observed at $H_{\rm 2}(\phi) \le H_{\rm
  c2}(\phi)$ also with four fold symmetry but out-of-phase modulation
(full triangles in Fig.\ \ref{Hcphi}). The polar angle is set to
$\theta=90^\circ$ throughout this subsection.

In Ref.\ \onlinecite{Wern02b} it has been shown that the
superconducting order parameter in Sr$_2$RuO$_4$ has two spatially
slightly anisotropic components described by the SO(2) vector
$\bOmega_{\rm f}$. The magnetic field couples strongest to the order
parameter component with dominant superconducting correlations
perpendicular to the field.

Generalizing Eq.\ (\ref{DeltaTofH}) to a two component vector
$\bDelta_T \sim \bOmega_{\rm f}$ the components $\nu=x,y$ of the order
parameter become  
\begin{equation}\label{Deltaxy}
\Delta_{T,\nu}(H,\phi) = \frac{\Delta_0}{\sqrt{2}}\ 
         \frac{H_{\rm c,\nu}(\phi,T) - H(\phi)}
                                           {H_{\rm c,\nu}(\phi,0)}\,. 
\end{equation}
Note that the superconducting correlations of the component
$\Delta_{T,x}$ are largest along $[110]$ while for $\Delta_{T,y}$
they are largest along $[\overline{1}10]$.\cite{Wern02b} The critical
fields can be parameterized into a homogeneous part and an alternating
part\cite{Sigr00} reflecting the two-fold symmetry of the order
parameter components.
\begin{equation}\label{Hcnu}
H_{{\rm c},{x \atop y}}(\phi,T) = H_{{\rm c},u}(T) 
                        + H_{{\rm c},a}(T)\ f(\pm\sin2\phi)
\end{equation}
The two components $H_{{\rm c},x}$ and $H_{{\rm c},y}$ only differ by
a phase shift $\phi \to \phi + \pi/2$ reflecting the symmetry relation
of $\Delta_{T,x}$ and $\Delta_{T,y}$. The total order parameter
$|\bDelta_T(H\to 0)|$ has the four-fold symmetry of the underlying
tetragonal lattice.\cite{Wern02b}

An appropriate choice for the anisotropy function is
\begin{equation}\label{deff}
f_\alpha(\sin2\phi) = \frac{\arcsin (\alpha \sin 2\phi)}
                           {\arcsin \alpha}\,
\end{equation}
that interpolates between a sinusoidal function for $\alpha\to 0$ and
a zig-zag function for $\alpha\to 1$. The fits in Fig.\ \ref{Hcphi}
for $H_{{\rm c},x}$ (dash-dotted line) and for $H_{{\rm c},y}$ (dashed
line) reveal that the experimental data are best described by
$\alpha\approx 1$. The parameters are determined to be $H_{{\rm
c},0}(T\to 0) = 1.44$ T and $H_{{\rm c},a}(T\to 0) = 0.05$ T. They are
consistent with the observations made by thermal conductivity
measurements.\cite{TNM+01,TSN+01} Note that the definition of
``$H_2$'' in Ref.\ \onlinecite{TNM+01} differs from the one made here.

The ratio $2H_{{\rm c},a}/H_{{\rm c},u}\approx 0.07$ is consistent
with the observed anisotropies in the thermal
conductivity in the superconducting phase.\cite{ITY+01} A more
detailed and quantitative comparison with the thermal conductivity
data requires a more involved analysis of the transport properties of
the model discussed here. It must include the symmetry breaking effect
of the temperature gradient.\cite{Wern02b} 

The fields $H_{{\rm c},x}$ and $H_{{\rm c},y}$ can be combined to
give the initial interpretation of the transitions in terms of 
\begin{equation}\label{Hc2ofHcxy}
H_{\rm c2}(\phi) = {\rm sup} 
           \left[H_{{\rm c},x}(\phi),H_{{\rm c},y}(\phi)\right]
\end{equation}
and 
\begin{equation}\label{Hpc2ofHcxy}
H_{\rm 2}(\phi) = {\rm inf} 
            \left[H_{{\rm c},x}(\phi),H_{{\rm c},y}(\phi)\right]\,,
\end{equation}
which then reflect the four-fold symmetry of the lattice.

The angular variation of the order parameter components in Eq.\
(\ref{Deltaxy}) with $\alpha=1$ is non-analytic at $\phi_n =
\frac{\pi}{4}(2n+1)$. This is consistent with the presence of quasi
one-dimensional correlations along the diagonals of the tetragonal
reciprocal lattice as predicted by the microscopic model for
Sr$_2$RuO$_4$.\cite{Wern02a} A more detailed analysis of how the quasi
one-dimensional correlations determine the order parameter would
require the study of the two-dimensional sine-Gordon
actions\cite{Wern02b} that determine the full Eliashberg\cite{Wern02e}
equation. This is not evident and must be left for future
studies. Experimentally it would be interesting to study the in-plane
anisotropy as shown in Fig.\ \ref{Hcphi} at various temperatures in
order to obtain the temperature dependence of $\alpha$.


\subsection{In-plane anisotropy of $H_{\rm p}$}\label{sectionHp}

The field $H_{\rm p}$ is defined by the upper edge of a shoulder
formed by an increase of the ac-susceptibility with increasing
reduced field $h$ as measured in Ref.\ \onlinecite{YAM+01}. The
experimental data are shown by the full diamonds in Fig.\ \ref{Hcphi}
and suggest a close relation of $H_{\rm p}$ to the critical order
parameter fields $H_{{\rm c},x}$ and $H_{{\rm c},y}$.

With an alternating field of 0.05 mT at frequencies of 700-1000 Hz the
ac-susceptibility measurements probe very low energy magnetic
excitations of the system.\cite{YAM+01} Within the framework of the 
underlying microscopic model the magnetic subsystem\cite{Wern02a} has an
energy scale of $\ov_{\rm eff} \sim 22$ K. Since typical magnetic fields
$H \sim 1$ T $\sim 0.7$ K $\ll \ov_{\rm eff}$ its contribution remains
essentially unaffected.\cite{Klum98} The spin triplet components
$\obOmega_{\rm s}$ of the order parameter is Zeeman split in a
magnetic field (see also Sec.\ \ref{sectionLTSH}). On the other hand,
interaction terms in the microscopic model couple the magnetic degrees
of freedom to the flavor degrees of freedom $\bOmega_{\rm f}$ of the
order parameter.\cite{Wern02b} The ac-susceptibility is thus an
indirect probe of the gapless SO(2) fluctuations of $\bOmega_{\rm f}$
in the superconducting state. 

The SO(2) fluctuations of $\bOmega_{\rm f}$ are well described by a
non-linear sigma model if the energy gap to the electronic (amplitude)
excitations is sufficiently large. Near the phase transition the
coupling to the amplitude excitations plays an important
role.\cite{Wern02b} The crossover between the two regions defines the
crossover gap $\Delta_c$. 

Only for $|\bDelta_T(H)| > \Delta_{\rm c}$ the gapless flavor
fluctuations\cite{Magneticquote} of the order parameter components can
be described by the 2+1 dimensional non-linear sigma model as
discussed in detail in Ref.\ \onlinecite{Wern02b}. For $|\bDelta_T(H)|
< \Delta_{\rm c}$ the system is in the ``effective saddle point''
regime where the amplitude mode energy cutoff plays a crucial role
and the internal degrees of freedom are integrated out. The field at
which the order parameter is equal to the crossover gap can be
identified as $H_{\rm p}$, i.e., $|\bDelta_T(H_{\rm p})| = \Delta_{\rm
  c}$. For smaller fields $H < H_{\rm p}$ one has $|\bDelta_T(H)| >
\Delta_{\rm c}$ and the system exhibits gapless flavor fluctuations
which increase\cite{MMN+00,YAM+01} the magnetic response through the
spin-flavor coupling.   

In order to derive the specific geometry and temperature dependence of
$H_{\rm p}$ within the framework of the present approach consider that
from Eqs.\ (\ref{Deltaxy}) -- (\ref{deff}) and $H_{{\rm c},u} \gg
H_{{\rm c},a}$ follows that the modulus of the order parameter is
\begin{equation}\label{absDelta}
\frac{|\bDelta_T(H,\phi)|^2}{\Delta_0^2} \approx 
\frac{\left[H_{{\rm c},u}(T) - H \right]^2 + 
               H^2_{{\rm c},a}(T) f^2_1(\sin2\phi)}
{H_{{\rm c},u}^2(0)}\,.
\end{equation}
The crossover field $H_{\rm p}(\phi)$ is defined by $|\bDelta_T
(H_{\rm p})| = \Delta_c$ and takes a simple form in the case $\phi=0$
or, equivalently, for $H_{{\rm c},a}(T) = 0$:
\begin{equation}\label{Hpphi0}
H_{\rm p}(0) = H_{{\rm c},u}(T) - 
    \frac{H_{{\rm c},u}(0)}{\Delta_0}\ \Delta_{\rm c}\,.
\end{equation}
The crossover field $H_{\rm p}$ is shifted by a temperature independent 
constant with respect to $H_{{\rm c},u}(T)$. 

For $H_{{\rm c},a}(T) \neq 0$ and $\phi \neq 0$ the SO(2) fluctuations 
between the order parameter components $\Delta_{T,x}$ and
$\Delta_{T,y}$ only are possible if the ratio 
\begin{equation}
r^2_\Delta = \frac{|\Delta^2_{T,x}-\Delta^2_{T,y}|}
               {|\bDelta_{T}|^2}
\end{equation}
is small enough, i.e., $r^2_\Delta < r^2_{\rm c}$. The theoretical form of
the angular dependent field then becomes
\begin{eqnarray}\label{Hpphi}
H_{\rm p}(\phi) &=& H_{{\rm c},u}(T) 
        - \frac{2 H_{{\rm c},a}(T)}{r_c^2}\ \left|f_1(\sin2\phi)\right|
\nonumber\\&&-\ 
    \sqrt{\frac{H^2_{{\rm c},u}(0)}{\Delta^2_0}\ \Delta^2_{\rm c} 
          - H^2_{{\rm c},a}(T) f^2_1(\sin2\phi)}\,.
\end{eqnarray}
The resulting fit to the experimental data at $T=0.06$ K is shown as
the dotted line in Fig.\ \ref{Hcphi} and yields the parameters
$\Delta_{\rm c} (T\to 0) = 0.4$ K and $r^2_{\rm c} = 0.64$.
The agreement with the experimental results is quite satisfactory.


\subsection{Temperature dependence}\label{sectionHa}

\begin{figure}[bt]
   \epsfxsize=0.48\textwidth
   \centerline{\epsffile{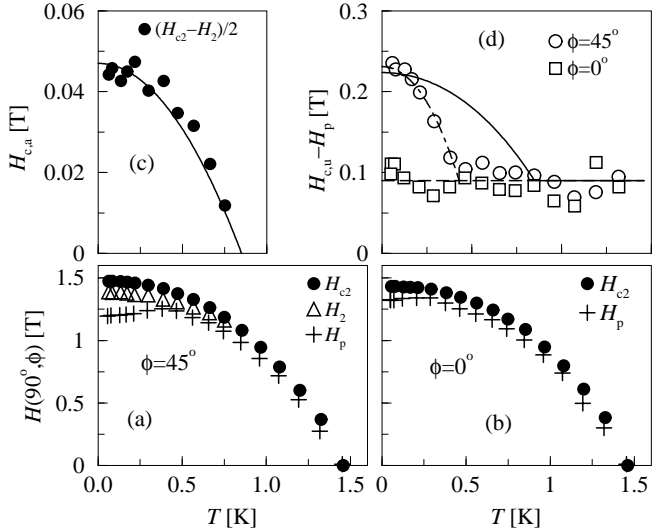}}
   \centerline{\parbox{\textwidth}{\caption{\label{H2andHp}
   \sl Panels (a) and (b) show the temperature dependence of the
   critical in-plane fields as derived from ac-susceptibility
   measurements\protect\cite{YAM+01} for $\phi=45^\circ$ and
   $\phi=0^\circ$, respectively. Panel (c) shows $H_{{\rm c},a} =
   (H_{\rm c2} - H_{\rm 2})/2$ together with a fit from Eq.\
   (\protect\ref{Hca(T)}). Panel (d) shows $H_{{\rm c},u}(T) -
   H_{\rm p}(\phi,T)$ from experimental data together with the respective
   fits from Eqs.\ (\protect\ref{Hpphi}) and (\protect\ref{Hca(T)})
   for $\phi = 45^\circ$ (circles, full line) and $\phi = 0^\circ$
   (squares, dashed line) with $H_{{\rm c},a}(0)=0.047$ T and $T_{{\rm
   c},a}=0.85$ K. The dash-dotted line is obtained using $H_{{\rm
   c},a}(0)=0.05$ T and $T_{{\rm c},a}=0.43$ K for $\phi =
   45^\circ$.}}}       
   \end{figure}

Figure \ref{H2andHp}(a) and (b) show the temperature dependence of the
fields $H_{\rm c2}$, $H_{\rm 2}$ and, $H_{\rm p}$ as derived from
ac-susceptibility measurements (Ref.\ \onlinecite{YAM+01}) for 
$\phi=45^\circ$ and $\phi=0^\circ$, respectively. The temperature
dependence of $H_{{\rm c},u}(T) = [H_{\rm c2}(T) + H_{\rm 2}(T)]/2$ is
essentially given by Eq.\ (\ref{HcofT}). From the difference 
$H_{\rm c2} - H_{\rm 2} = 2H_{{\rm c},a}$ at $\phi=45^\circ$ the
temperature dependence of the anisotropic field component can be
extracted as 
\begin{equation}\label{Hca(T)}
H_{{\rm c},a}(T)\big|_{T \le T_{{\rm c},a}}
     \approx H_{{\rm c},a}(0)\ 
                       \left(1-\frac{T^2}{T_{{\rm c},a}^2}\right)
\end{equation}
with $H_{{\rm c},a}(0)=0.047$ T, $T_{{\rm c},a}=0.85$ K and $H_{{\rm 
c},a}(T)|_{T > T_{{\rm c},a}} = 0$. The experimental data
(symbols) and the fit (full line) are shown in  Fig.\
\ref{H2andHp}(c). From Eqs.\ (\ref{Deltaxy}) and (\ref{Hcnu}) 
follows that the difference $\Delta_{T,x}-\Delta_{T,y}\approx 0.2$ K
for $\phi=45^\circ$ ($H\le H_{\rm 2}$). Consequently the vanishing of
$H_{{\rm c},a}(T)$ for $T>T_{{\rm c},a}$ can be associated with
thermal fluctuations obscuring the difference between the order
parameter components $\Delta_{T,x}$ and $\Delta_{T,y}$.

The symbols in Fig.\ \ref{H2andHp}(d) show the data for $H_{{\rm
c},u}(T) - H_{\rm p}(\phi,T)$ as extracted from the experimental
data for $\phi = 45^\circ$ (circles) and $\phi = 0^\circ$
(squares). The broken line is the temperature independent result from
Eq.\ (\ref{Hpphi}) for $\phi = 0^\circ$  with $\Delta_{\rm c} = 0.3$
K in excellent agreement with experiment. The discrepancy between the
values extracted here [$\Delta_{\rm c} = 0.3$ K and $H_{{\rm c},a}(0)
= 0.047$ T] and the values extracted from the angular dependence shown
in Fig.\ \ref{Hcphi} [$\Delta_{\rm c} = 0.4$ K and $H_{{\rm c},a}(0) =
0.05$ T] can in part be attributed to experimental uncertainties and in
part to the corrections to the theory for $T<0.4$ K as anticipated in
Sec.\ \ref{sectionFL}.  

For $\phi = 45^\circ$ the experimental data [circles in Fig.\
\ref{H2andHp}(d)] are only in qualitative agreement with the fit form
Eq.\ (\ref{Hpphi}) [full line, $\Delta_{\rm c} = 0.3$ K, $r^2_{\rm c}
= 0.64$, $H_{{\rm c},a}(0)=0.047$ T, $T_{{\rm c},a}=0.85$ K] for $0.2\
{\rm K} < T < 0.8\ {\rm K}$. This might indicate that there are
corrections to the theory. On the other hand the read-off from the
ac-susceptibility data of $H_{\rm p}$ for $T < 0.6\ {\rm K}$ and of
$H_{\rm 2}$ for $T > 0.6\ {\rm K}$ is not unambiguous.\cite{YAM+01} To
match the experimental data in Fig.\ \ref{H2andHp}(d) parameters
$H_{{\rm c},a}(0)=0.05$ T and $T_{{\rm c},a}=0.43$ K  are adequate
[dash-dotted line in Fig.\ \ref{H2andHp}(d)]. A more complex
temperature dependence with a small $H_{{\rm c},a}(T) < 0$ for $T >
0.7 T_{\rm c}$ has been proposed in Refs.\
\onlinecite{Sigr00,MMN+00}. An angular analysis of the
ac-susceptibility as shown in Fig.\ \ref{Hcphi} at temperatures $0.2\
{\rm K} < T < 0.8\ {\rm K}$ would be helpful. Complimentary data from
specific heat measurements as a function of the in-plane field
direction are also desirable. 

The crossover or ``fluctuation field'' $H_{\rm p}$ marks the
appearance of quasi two-dimensional fluctuations as described by 
the non-linear sigma model. In that sense $H_{\rm p}$ marks a
phenomenon similar to a Kosterlitz-Thouless transition.\cite{Tsve95} 


\subsection{Interpretation of $\Delta_{\rm c}$ in terms of the
  microscopic model}\label{sectionDeltac}  

For reduced temperatures $t \le t_{\rm c} \approx 0.05$ the order
parameter fluctuations are in the ``effective saddle point'' regime as
defined in Ref.\ \onlinecite{Wern02b}. In the effective  saddle point
regime the superconducting gap defines a cutoff in the spectrum of
the gapless SO(2) fluctuations of the internal flavor degrees of
freedom of the order parameter.\cite{Magneticquote} Gapless
fluctuations of the order parameter components as described by the 2+1
dimensional non-linear sigma models without an energy cutoff only
exist for $t (H\to 0) > t_{\rm c}$. 

The analogy between the reduced field $h$ and the reduced temperature
$t$ (Sec.\ \ref{sectionFL}) implies that for $h(T\to0) \le h_{\rm c}
\approx 0.05$ the system also is in the effective  saddle point
regime. The value of $\Delta_{\rm c} \approx 0.4$ K (Sec.\
\ref{sectionHp}) gives together with Eq.\ (\ref{DeltaTofH}) $h_{\rm
c}(T\to0) \approx 0.08$. Considering the experimental uncertainties
and the simplicity of the approach this agreement is quite
satisfactory. Equation (\ref{DeltaTofH}) can be rewritten as
\begin{equation}\label{hcofDeltac}
h_{\rm c}(T) = \frac{\Delta_{\rm c}}{\Delta_0}\ 
                     \frac{H_{\rm cm}(0)}{H_{\rm cm}(T)}
\end{equation}
and reveals the temperature dependence of $h_{\rm c}(T)$. 

Similarly follows from Eq.\ (\ref{DeltaHofT}) that with $\Delta_{\rm
c} \approx 0.3$ K (Sec.\ \ref{sectionHa}) $t_{\rm c}(H\to0) \approx
0.06$. Again, the agreement with the proposed value of $t_{\rm
c} \approx 0.05$ underlines the consistency of the microscopic
model in Refs.\ \onlinecite{Wern02a,Wern02b} and the phenomenological
approach herein.

The analysis discussed herein suggests that the maximum in the specific
heat below the superconducting transition coincides with $H_{\rm
p}$. Probing the specific heat dependent on the in-plane field
direction experimentally would be an appropriate test of this
prediction.  


\subsection{Out-of-plane spin-flip field $H_{\perp}$ and out-of-plane
  anisotropy of $H_{{\rm c},a}$}\label{sectionout}  

At low temperatures and for small out-of-plane fields $H(0,\phi) <
H_{\perp} \| H(0,\phi)$ the thermodynamic and transport properties of
the system are basically unaffected. The value $H_{\perp} \approx
0.01$ T can be determined both from the specific heat\cite{NMM00} as
well as from thermal conductivity\cite{ITY+01,STM+02}
measurements. Obviously there is a small confinement of the magnetic
moment of the Cooper pairs to the $x$-$y$ plane that becomes apparent
when fluctuations are frozen out. The spin-flip field $H_{\perp}$ is
found to be temperature independent as long as $H_{\rm c2}(0,\phi,T) > 
H_{\perp}$\cite{ITY+01,STM+02} and shows a temperature dependent
hysteresis.\cite{STM+02} 

The dependence of the field $H_{\rm 2}(\theta,45^\circ)$ on the
polar angle $\theta$ as observed in Ref.\ \onlinecite{YAM+01} is
closely related to the out-of-plane spin-flip field. For $\theta <
89.5^\circ$ the out-of-plane component of the magnetic field
$\hat{\bz}\bH$ is larger than the spin-flip field, i.e., $\hat{\bz}\bH
> H_\perp$. For $\theta < 89.5^\circ$ the order parameter has an
out-of-plane spin component which couples homogeneously to both
spatial components $\Delta_{T,x}(H)$ and $\Delta_{T,y}(H)$. $H_{{\rm 
c},a}$ vanishes and $H_{\rm 2} = H_{\rm c2}$. The transition from
$H_{{\rm c},a}(\theta > 89.5^\circ)\neq 0$ to $H_{{\rm c},a}(\theta <
89.5^\circ) = 0$ should be of first order for $T \le 0.02$ K.


\section{Summary}

Based on the microscopic model introduced in Refs.\
\onlinecite{Wern02a} and \onlinecite{Wern02b} the observed specific
heat and  ac-susceptibility data in the superconducting phase in
Sr$_2$RuO$_4$ with applied magnetic fields have been described
consistently.

\subsection{Conclusions}

(i) The temperature dependence of the upper critical field is
satisfactorily described by a phenomenological formula similar to the
BCS results [Eq.\ (\ref{HcofT})].

(ii) The dependence of the upper critical field on the out-of-plane
angle of the field direction is excellently reproduced by the
Landau-Ginzburg anisotropic effective mass approximation if the
enhanced in-plane coupling is included [Eq.\ (\ref{Hcofthetaani})].   

(iii) The specific heat below the superconducting phase transition
increases linearly with decreasing temperature as predicted in the
presence of quasi two-dimensional gapless fluctuations of the order
parameter. The observed reduction of the temperature slope with
increasing magnetic field is consistent with the reduced magnetic
degrees of freedom. The resulting linear dependence of the specific
heat on the magnetic field just below the critical field is
temperature independent and in quantitative agreement with
experiment. 

(iv) The two spatially anisotropic components of the order parameters
couple differently to the applied magnetic field depending on its
orientation. The observed angular dependence of the resulting two
critical fields leads to conclude that each of the components of the
order parameter has a spatial anisotropy of $\sim 7$ \% for $H\to 0$. 

(v) The non-analytic angular variation of the in-plane upper
critical fields supports the presence of quasi one-dimensional
correlations along the diagonals of the basal plane of the unit cell
as predicted by the underlying microscopic model.   

(vi) The order parameter fluctuations are described for fields $H \ll 
H_{\rm p}$ by the non-linear sigma model while for $H_{\rm p} < H <
H_{\rm c2}$ they are qualitatively altered by the coupling to the
amplitude fluctuations. Both the angular and temperature dependence of
the fluctuation field $H_{\rm p}$ are described consistently with
ac-susceptibility measurements. 

(vii) For small temperatures $T<0.1$ K the spin-gap field $H_{\rm sg}
\approx 0.12$ T has been determined. For fields $H < H_{\rm sg}$ the
spectrum of the internal magnetic degrees of freedom of the order
parameter $\obOmega_{\rm s}$ has a gap mediated by spin-flavor
coupling. For $H = 0$ the gap (Leggett frequency) has been estimated
as $\Omega_A \sim 0.08$ K.

\subsection{Proposed experiments}

For further clarification of the description of the low temperature
electronic properties of Sr$_2$RuO$_4$ the following additional
experiments would be useful.
 
(i) An angular analysis of the ac-susceptibility at temperatures $0.2\
{\rm K} < T < 0.8\ {\rm K}$ would be helpful to clarify if there are
corrections to the theory concerning the angular dependence of the
field $H_{\rm p}$ in that temperature range. 

(ii) Complimentary data from specific heat measurements as a function
of the in-plane field direction are desirable to test if the maximum
of the specific heat coincides with $H_{\rm p}$.

(iii) Specific heat measurements in the range of $0<T<0.4$ K and
$1.4<H<1.5$ T are desirable to study details of the slope of the
linear field and temperature dependence of the specific heat near the
transition. Possible corrections to the theory in that parameter range
can then be investigated.

(iv)  Excess current measurements provide a quantitative test of the
linear dependence of the order parameter on the temperature and the
magnetic field as predicted in (\ref{DeltaHofT}) and
(\ref{DeltaTofH}).  


\subsection{Critique and outlook}

The aim of the presented analysis of the properties of Sr$_2$RuO$_4$
in magnetic fields is to consistently describe a large number of
experimental results within a minimal model. To this end two
assumptions where made. The first is the separation of the magnetic
and flavor degrees of freedom  as imposed by the underlying
microscopic model discussed in Ref.\ \onlinecite{Wern02a} and
\onlinecite{Wern02b}. Secondly it is assumed that the magnetic field
dependence of the order parameter can be described by simple relations
similar to the temperature dependence as given in Eqs.\
(\ref{DeltaHofT}) and (\ref{DeltaTofH}). The qualitative and
quantitative agreement of the specific heat data (Secs.\
\ref{sectionFL} and \ref{sectionLTSH}) and the geometry of the
critical fields (Secs.\ \ref{sectionHc2}, \ref{sectionHp} and
\ref{sectionHa}) underline the validity of the approach.  

Based on these results a more detailed study of the interplay of the
magnetic and flavor degrees of freedom together with the vortex
lattice via a full Landau Ginzburg analysis appears desirable. Similar 
approaches were performed earlier for a two-component $p$-wave order
parameter\cite{Sigr00,Agte01,Baba02} and $^3$He.\cite{VK83,Volo92}


\section*{Acknowledgments}

I thank M.\ Eschrig, J.\ Kroha, and F.\ Laube for discussions.
The work was supported by DOE contract number DE-AC02-98CH10886 and
the Center for Functional Nano\-struc\-tures at the University of
Karlsruhe.


\begin{thebibliography}{10}
\bibitem{Magneticquote}
The fluctuations of the internal magnetic degrees of freedom of the order
  parameter are assumed to be frozen out by the magnetic field, c.f.\ Sec.\
  \ref{sectionLTSH}.
\end{thebibliography}
\end{document}